%% file: main.tex
\documentclass[runningheads]{llncs}

\usepackage[utf8]{inputenc}
\usepackage[T1]{fontenc}
\usepackage{graphicx}
\usepackage{geometry}
\usepackage{amsmath}
\usepackage{mathtools}
\usepackage[dvipsnames]{xcolor}
\usepackage{float}
\usepackage{algorithm}
\usepackage[noend]{algpseudocode}
\usepackage{xspace}
\usepackage[hidelinks]{hyperref}
\usepackage{todonotes}
\usepackage{csquotes}
\usepackage{fullpage}

\newcommand{\algname}{{\sc Saber}\xspace}
\graphicspath{ {./images/} }
\newcommand{\junk}[1]{}

\title{Pairwise sequence alignment with block and character edit operations}

\author{Ahmet Cemal Alıcıoğlu\inst{1,2} \and
Can Alkan\inst{1}
}%
\authorrunning{A.C. Alıcıoğlu et al.}
\institute{Dept of Computer Engineering, Bilkent University, Ankara 06800, Turkey \and
Dept of Computer Science, University of Freiburg, Freiburg im Breisgau 79110, Germany \\
\email{alicioga@informatik.uni-freiburg.de, calkan@cs.bilkent.edu.tr } \\
}

\date{\today}
\begin{document}

\maketitle

\begin{abstract}
    Pairwise sequence comparison is one of the most fundamental problems in string Pairwise sequence comparison is one of the most fundamental problems in string processing. The most common metric to quantify the similarity between sequences $S$ and $T$ is \textit{edit distance}, which corresponds to the number of characters that need substituted, deleted from, or inserted into $S$ to generate $T$. However, if larger rearrangements are permitted, fewer edit operations may be sufficient for some string pairs to transform one string to another.  \textit{Block edit distance} refers to such changes in substring level (i.e., \textit{blocks}) that  ``penalizes'' entire block removals, insertions, copies, and reversals with the same cost as single-character edits. Most studies to calculate block edit distance to date aimed only to characterize the distance value for applications in sequence nearest neighbor search without reporting the full alignment details. Although a few tools try to solve block edit distance for genomic sequences, they have limited functionality and are no longer maintained.
    Here, we present \algname, an algorithm to solve block edit distance that supports block deletions, moves, and reversals in addition to the classical single-character edit operations. Our algorithm runs in $O(m^2\cdot n\cdot \ell_{\textit{range}})$ time for $|S|=m, |T|=n$ and the permitted block size range of $\ell_{\textit{range}}$; and can report all breakpoints for the block operations. We also provide an implementation of \algname currently optimized for genomic sequences, although the algorithm can theoretically be used for any alphabet. 
    \\
    \hspace*{0.5cm} \algname is available at \href{http://github.com/BilkentCompGen/saber}{http://github.com/BilkentCompGen/saber}
\end{abstract}

\clearpage

\section{Introduction}

\input{intro}

\section{Preliminaries}
\label{sec:prelim}
\input{preliminaries}

\section{Methods}
\label{sec:methods}
\input{methods}

\section{Results}
\label{sec:results}
\input{results}

\section{Author contributions statement}

A.C.C. developed and tested the \algname. A.C.C. and C.A. wrote the manuscript.

%USE THE BELOW OPTIONS IN CASE YOU NEED AUTHOR YEAR FORMAT.

\bibliographystyle{splncs04}
\bibliography{calkan}

\input{supp}

\end{document}

%% file: intro.tex
Sequence comparison is a well-studied problem in combinatorial pattern matching. Several proposed algorithms allow for single-character substitutions, short insertions, and deletions (indels)~\cite{Levenshtein1966,Medvedev2023}. Larger events can also contribute to the ``distance'' between two sequences, including block edits such as deletions, insertions, reversals, and block moves~\cite{Batu2005,Bourdaillet2007,Muthukrishnan2004,Shapira2007}. Smith-Waterman algorithm~\cite{Smith1981} has been the standard for local similarity detection, especially for biological sequences; however, it cannot detect multiple block rearrangements.

The classical definition of edit distance refers to the number of substitutions, insertions, and deletions to transform one string into another. Benson extended the substitution/indel (SI) model to include tandem duplications, hence introducing the \textit{DSI model}~\cite{Benson1997} and provided an exact algorithm that can be computed in $O(m^4)$ time, and a faster heuristic that runs in $O(m^2)$ time. Several other studies focused on block moves only~\cite{Bourdaillet2007,Shapira2007}.

A more generalized model uses \textit{block edit distance} (or \textit{edit distance under block operations})~\cite{Lopresti1997,Sahinalp2008}, which is defined as the number of such changes in blocks (i.e., substrings) in the form of removals (i.e., deletions), additions (i.e., insertions), reversals, and moves (i.e., translocations/transpositions). Denoted as $BED(S, T)$, the block edit distance problem, including single-character edits, is shown to be NP-Hard; hence no efficient exact algorithms exist~\cite{Ganczorz2018,Gonen2019,Lopresti1997,Sahinalp2008}.

Previous attempts at solving the block edit distance problem mainly focused on computing only the distance (i.e., no traceback) for applications in document exchange problem~\cite{Cormode2000} and sequence nearest neighbor problem~\cite{Muthukrishnan2000}. Owing to the computational complexity of the general problem, these studies presented approximate algorithms~\cite{Batu2005,Sahinalp2008,Gonen2019} or focused on variants with additional constraints that allowed for polynomial time solutions~\cite{Lopresti1997}.

%%%%%%%%%%%%%%%%%%%%%%%%%%%
% transition into genomics...

The block edit distance problem is relevant to the bioinformatics field. This is because mutational processes drive genome evolution in the form of single character substitutions~\cite{1000GP2015}, single-to-many character insertions and deletions~\cite{Mills2006}, and large-scale rearrangements such as inversions, duplications, and translocations~\cite{Alkan2011}, which correspond to block edit distance models using only the DNA alphabet (i.e., $\Sigma=\{A, C, G, T\}$). Due to both the computational complexity of block edit distance calculation and the fact that the genomic sequences are not typically generated in full and accurately (i.e., assembled), such rearrangements are only discovered through analyzing read mapping properties~\cite{Alkan2011,Chaisson2019,Ho2019} or comparing fragmented draft assemblies~\cite{Heller2020} in practice. 

Several algorithms have also been proposed that calculate pairwise alignments of genomic sequences with rearrangement discovery support, such as Shuffle-LAGAN~\cite{Brudno2003}, Mauve~\cite{Darling2004}, and GR-Aligner~\cite{Chu2009} with some limitations. For example, neither Shuffle-LAGAN nor Mauve report detailed alignments with complete information on rearrangement breakpoints and can find only block copies, block moves, and block reversals. On the other hand, GR-Aligner can report the alignments with rearrangements, but it focuses only on simple block reversals and block moves, and it is no longer maintained.

Extension of the block edit distance problem to discover genomic inversions was proposed~\cite{Hannenhalli1995}, and then generalized as the HP model for genomic rearrangements~\cite{Hannenhalli1995a}. Later, the double cut and join (DCJ) model was proposed as a general framework for rearrangements~\cite{Bergeron2006,Yancopoulos2005}. However, both HP and DCJ models require prior knowledge of conserved genomic markers, e.g., genes. Additionally, these markers have to be unique in each genome, which renders the models impractical due to the duplications in real genomic sequences. More recently, Bohnenkämper et al. introduced an exact algorithm to solve the genomic rearrangement problem with duplicated markers~\cite{Bohnenkaemper2020}.

%since complex rearrangements (i.e., multiple edits at the same location) make obtaining high accuracy in alignments harder.

Here, we introduce \algname (Sequence Alignment using Block Edits and Rearrangements), a pairwise sequence alignment algorithm under block edit distance models. \algname extends the models and algorithms provided by Lopresti \& Tomkins~\cite{Lopresti1997} to report the rearrangement breakpoints and single character changes. Furthermore, unlike previous studies~\cite{Hannenhalli1995,Hannenhalli1995a,Bergeron2006,Yancopoulos2005,Bohnenkaemper2020}, \algname does not require a set of genomic markers and operates directly on sequences. This paper focuses on genomic sequence alignments (i.e., using the DNA alphabet); however, our algorithms are not theoretically limited to a specific alphabet. As detailed in Section~\ref{sec:prelim}, our algorithms are based on a variant of the block edit distance model that allows for polynomial time solution~\cite{Lopresti1997}. \algname runs in  $O(m^2\cdot n\cdot \ell_{\textit{range}})$ time, where $m,n$ are the sequence lengths and $\ell_{\textit{range}}$ is the
range of permitted block sizes (i.e., $\ell_{\textit{range}}=\ell_{\textit{max}}-\ell_{\textit{min}}+1$).

%% file: preliminaries.tex
Given a source string $S = s_1s_2...s_m$ and a target string $T = t_1t_2...t_n$, we define the \textit{block edit distance} between $S$ and $T$, denoted as $BED(S, T)$, as the minimum number of edit operations to transform $S$ into $T$ including both single character changes (i.e., substitutions and insertions/deletions) and substring changes (i.e., block edits). We define a \textit{block} as a substring within $S$ or $T$. 
Operations allowed for the block edits are:

%Figure for each operation individually
\begin{figure*}[!htb]
    \centering
    \includegraphics[width=0.8\textwidth]{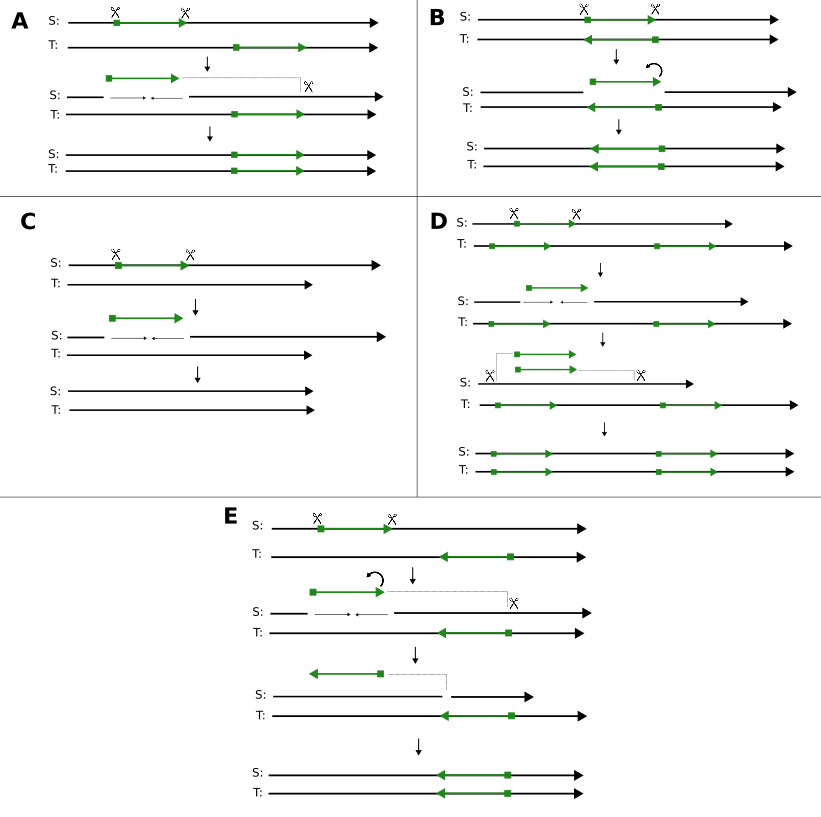}
    \caption{Block edit operations. A) block move, B) block reversal, C) block deletion, D) block copy, E) block move with a reversal (i.e., inverted block move). The current version of \algname supports operations depicted in A, B, C, and E. }
    \label{fig:operations}
\end{figure*}

\begin{figure}[!htb]
    \centering
    \includegraphics[width=0.45\textwidth]{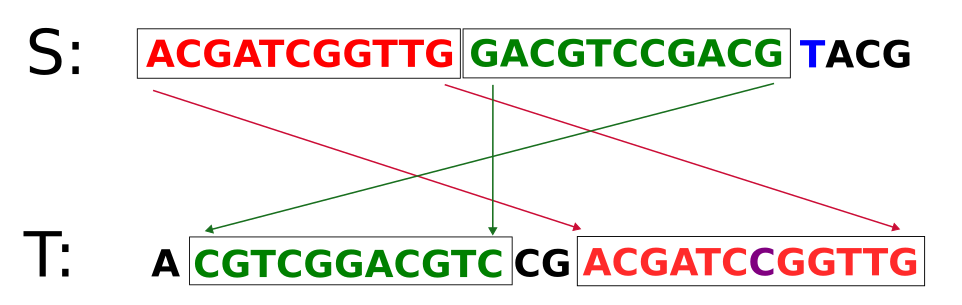}
    \caption{Pairwise alignment of sequences under block edit distance model. Here, we show one block move (red, cost=1) with a single character insertion (C in purple, cost=1), one inverted move (green, total cost=2), and one single character deletion (T in blue, cost=1). The total block edit distance is, therefore, 5.}
    \label{fig:example}
\end{figure}

\begin{enumerate}
    \item  \textbf{Single character edits,} specifically \textit{insertions, deletions} (i.e., \textit{indels}) and \textit{substitutions.} 
    \item  \textbf{Block moves,} where a contiguous substring (block) $b$ of length $\ell > 1$ from string $S$ is removed from its original position and inserted at a new position within $S$ (Figure~\ref{fig:operations}A). 
    \item  \textbf{Block reversals,} where the character order of characters in a block $b$ is reversed, and each character in this block is replaced with its complementary character (e.g., $A \leftrightarrow T, C \leftrightarrow G$ for the DNA alphabet; Figure~\ref{fig:operations}B). 
    \item  \textbf{Block deletions,} where a block $b$ is removed entirely from $S$ (Figure~\ref{fig:operations}C).
    \item  \textbf{Block copies,}  where a block $b$ is duplicated within $S$, creating an additional copy of $b$ in another starting position (Figure~\ref{fig:operations}D). The current version of \algname does not support block copies.
    \item  \textbf{Inverted block moves,} where a block is relocated from one place to another in an inverted orientation while also replacing the characters with their complementary characters. This operation combines the actions of a block move and a block reversal. (Figure~\ref{fig:operations}E).
\end{enumerate}
An additional particular constraint is that the block operations must not overlap, i.e., no character in the sequence can be involved in more than one block operation. The only exception to this rule is encapsulated in the \textit{inverted block move} operation. This particular operation is distinct in inherently involving a \textit{block move} and a \textit{block reversal}. The character operations may be used alongside any other operation. 
An example set of block and single character edit operations is shown in Figure~\ref{fig:example}. 

% component of rearrangement distance, since the edit distance component is well studied~\cite{Levenshtein1966,Sosic2017}.

%for now: The block edit distance problem here should not be confused with the block alignment defined for the Four Russians Speedup of the edit distance~\cite{Brubach2018}. In this problem, blocks can be rearranged, and the blocks are not known beforehand.

\subsection{Block Edit Models}
Since the domain of the approximate solution described in~\cite{Batu2005} is very limited, it cannot be used to detect similarities and only reports the block edit distance. Instead, we build the foundation of the block edit distance approximation in this paper based on the ideas presented earlier~\cite{Lopresti1997}. 

Lopresti \& Tomkins define the block edit distance problem as identifying the optimal set of block matches between two strings, aiming to minimize the cumulative distances of all block matches~\cite{Lopresti1997}. They also analyzed potential variations of this problem and discovered that for some combinations of restrictions on the sequences, this optimization problem can be solved in polynomial time using dynamic programming. The first restriction is \textit{coverage} (C), which means that blocks must comprehensively span every character in the string. The absence of the coverage restriction (denoted $\overline{C}$) would benefit the alignment of genomic sequences by permitting operations on characters outside the predefined blocks. The second restriction is \textit{disjointness} (D), which means that blocks should not overlap in terms of their position in the string.  (similarly, $\overline{D}$ denotes the absence of disjointness). For the block edit distance problem to be solvable in polynomial time, the target sequence must not have any restriction on coverage or disjointness, meaning the blocks selected from the target sequence may not cover every character in the string and may overlap.

The best problem model for applications to genomic sequences would then be the $\overline{C}D-\overline{C}D$ model (for $S$ and $T$, respectively)\footnote{No coverage restriction for either $S$ or $T$, disjointness restriction is applied for both $S$ and $T$.}. This ensures the disjointness for both sequences and the absence of the coverage restriction. However, this problem model is proven to be NP-complete~\cite {Lopresti1997}. Through relaxing requirements, the authors present an exact solution to the $\overline{C}D-\overline{CD}$ model \footnote{No coverage restriction for either $S$ or $T$, disjointness restriction is applied for $S$ but not for $T$.} that runs in $O(m^2n)$ time. However, this solution may not be directly applied to genomic sequences for the following reasons:
\begin{itemize}
    \item This model only considers the block moves and character edits within the moved blocks. It does not consider block removal, copy, reverse operations, or character edits outside the moved blocks, which can be observed in evolutionary processes.
    \item In the block edit distance problem for genomic sequences, blocks from both sequences must be disjoint because two overlapping block moves are relatively rare.
    \item The DNA alphabet is much smaller than the generalized alphabets used in~\cite{Lopresti1997}. 
    Therefore, a size restriction on the blocks is necessary.  
    \item There is no such thing as a ``negative distance'' in genomic sequence comparisons, making the absence of coverage in the source sequence more difficult to formulate.
\end{itemize}

We have developed a heuristic solution to block edit distance using a modified version of the solution given in~\cite{Lopresti1997}.

%% file: methods.tex
Here, we present \algname as a heuristic approach for solving the block edit distance problem with traceback, using a modified version of the solution to $\overline{C}D-\overline{CD}$ model, resolving the previously identified issues associated with the original methodology.

Similar to the algorithm described in~\cite{Lopresti1997}, \algname includes computing two arrays named $W$ and $N$ using the following two functions. \texttt{Compute-W} procedure matches every ``valid'' block in $S$ with the best matching block from $T$. The second function, \texttt{Construct-N}, uses the $W$ array to heuristically calculate the best set of block and character operations to transform $S$ into $T$. Below, we provide more details about the computations of these functions.

\subsubsection{Computing W.}
The \texttt{Compute-W} function aims to characterize the best matching between all possible blocks in $S$ and $T$.  $W$ is formulated as a two-dimensional matrix, and $W(i, j)$ holds the best score for block (i.e., substring) from $S$ starting from index $i$ and of length $j$ (i.e., $S[i..i+j-1]$). Block matches with low scores are disregarded based on a user-specified error rate to avoid unnecessary block operations. 

%If the input is long reads generated by a sequencing technology such as PacBio or ONT, the error rate should be selected depending on the error profile of 

Let $\mathcal{C}$ be an indicator variable for a potential block match with ``some'' character edits, defined as: 
$$
\mathcal{C}:\ \ \text{match\_dist} \le \Big \lceil error\_rate \times \frac{l_1 + l_2}{2} \Big \rceil
$$
where $l_1$ and $l_2$  are the lengths of the matching blocks.
For example, with an error rate of 10\% and blocks of length 50, the algorithm will allow at most five character edit operations between the blocks.

Additionally, to consider a \textit{Block Reverse} operation within a \textit{Block Move}, we use the  following distance function instead of edit distance:
$$
dist_{\text{BR}}(X, Y) = min\{ed(X, Y), ed(X, Y_{rc}) + C_{\text{BR}}\}
$$
where $ed$ is the edit distance~\cite{Levenshtein1966}, $Y_{rc}$ is the reverse complement of $Y$, and $C_{\text{BR}}$ is the cost of reversing a block.

With these definitions, building the $W$ array using the formulae below is now possible.

$$
best\_dist_{i,j} = \underset{0 \le k \le n}{\underset{\ell_{\textit{min}}\le \ell\le \ell_{\textit{max}}}{min}}\{dist_{\text{BR}}(S[i..i+j-1], T[k..k+\ell-1])\} 
$$
$$
W(i,j-\ell_{\textit{min}}) = \begin{cases} best\_dist_{i,j} & \text{if } best\_dist_{i,j} \text{ satisfies } \mathcal{C} \\+\infty &\text{otherwise}\end{cases}
$$
where $\ell_{\textit{min}}$ and $\ell_{\textit{max}}$ are the minimum and maximum block lengths specified as parameters to \algname. The range of block length ($\ell_{\textit{min}}, \ell_{\textit{max}}$) should be given as the length of expected rearrangements that the user aims to detect.

Note that $W$ only stores distances to the best matches. \algname uses another array along with $W$ to keep track of matches and determine whether they represent a \textit{block move} or an \textit{inverted block move}.

\subsubsection{Constructing N.}
To obtain the final block edit distance using $W$, we define array $N$ of size $m$, which holds the best set of block operations by storing the operation performed at every index of the source sequence $S$. We consider the following operations at each index: 
\begin{itemize}
\item Skip the character.
\item Use a block match of length $\ell_{\textit{min}}\le \ell\le \ell_{\textit{max}}$ as block move 
by querying the $W$ array.
\item Remove the block of length $\ell_{\textit{min}}\le \ell\le \ell_{\textit{max}}$  as a block remove.
\end{itemize}

More formally, given that $W$ was already calculated, the problem in \texttt{Construct-N} is to fill the array $N$ with possible operations to minimize the final block edit distance (calculated using Algorithm~\ref{alg:bes}). An element in $N$ can have the following values:
\begin{itemize}
\item $N(i) = j,$ indicating block move of length $j$, $S_i...S_{i+j-1}$ according to $W(i, j-\ell_{\textit{min}})$, where $\ell_{\textit{min}} \le j \le \ell_{\textit{max}}$.
\item $N(i) = -j$, indicating block remove of length $j$, removes all characters $S_i...S_{i+j-1}$, where $\ell_{\textit{min}} \le j \le \ell_{\textit{max}}$.
\item $N(i) = 0$, indicating skipping the character $S_i$ at this index.
\end{itemize}

The first $\ell_{\textit{min}}$ indices must be skipped because no move or remove operations are viable within this range. Another constraint is that no block operations may overlap on either sequence $S$ or $T$.

\begin{figure}[!ht]
    \centering
    \includegraphics[width=0.42\textwidth]{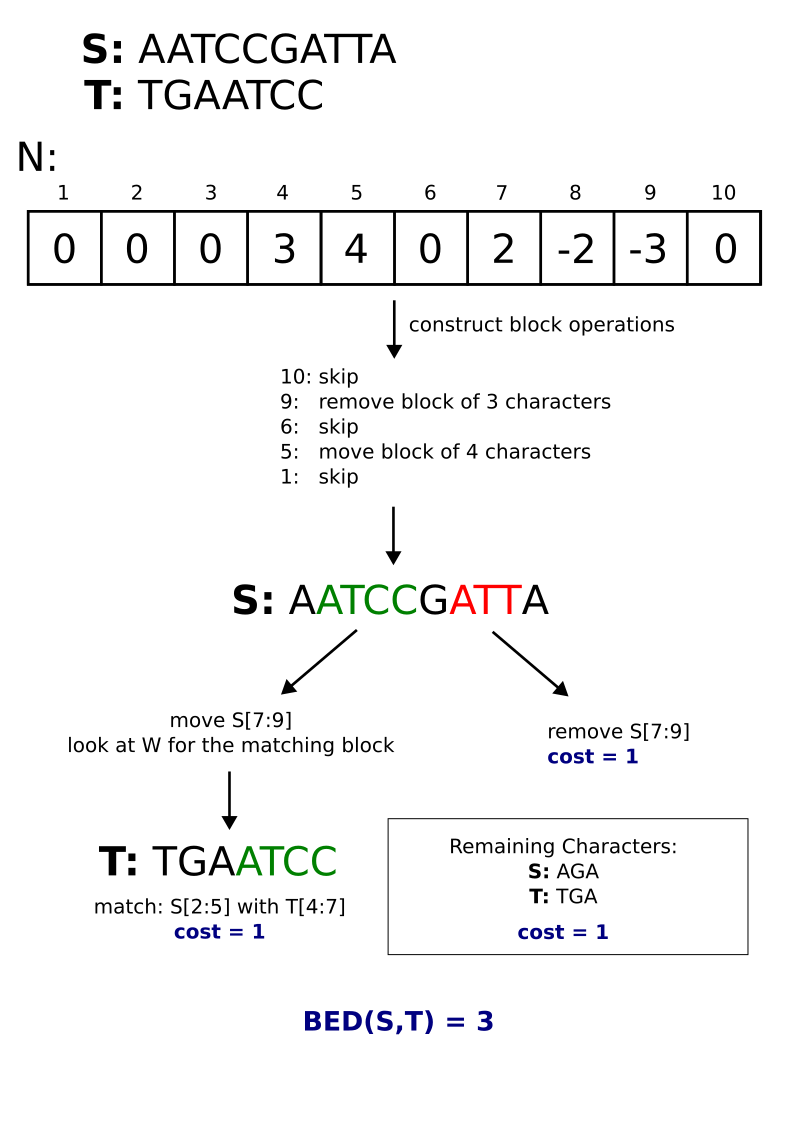}
    \caption{Example calculation of \texttt{Block-Edit-Distance} (BED) from a given $N$ array.}
    \label{fig:calculate_n}
\end{figure}

Note that there exist ${(2\times \ell_{\textit{range}} + 1)}^{m-\ell_{\textit{min}}}$ possible configurations (i.e., settings) of $N$. The primary goal of \texttt{Construct-N} function is to discover the configuration that calculates the block edit distance (BED). Our heuristic solution achieves this by iteratively selecting the optimal operation for each index within the array $N$. This selection is based on assessing how each potential operation impacts the local block edit distance, ultimately choosing the operation that minimizes it locally. The calculation of block edit distance involves retracing the $N$ array in reverse to identify all block operations. After identifying these operations, we calculate the total cost associated with each block operation.  This procedure is detailed in Algorithm~\ref{alg:bes} and illustrated in Figure~\ref{fig:calculate_n}.

\begin{algorithm}[!htb]
\caption{Block Edit Distance}\label{alg:bes}
\begin{algorithmic}[1]
\Procedure{Block-Edit-Distance}{$S, T, W, N$}
\State Reading backwards from $N$, identify all the block operations involved.
\State $S_\textit{rem}$ $\gets $ remaining characters in $S$ that are not covered in any of the block operations.
\State $T_\textit{rem}$ $\gets$ remaining characters in $T$ that are not covered in any block operations.
\State $C_\textit{char}$ $\gets$ edit\_distance($S_\textit{rem}$, $T_\textit{rem}$)
\State Initialize $C_\textit{block}$ to $0$
\For{\textbf{each} block operation identified}
    \State $C_\textit{block} \gets C_\textit{block}~+$ cost of this block operation
\EndFor
\State
\Return $C_\textit{char} ~ + ~ C_\textit{block}$
\EndProcedure
\end{algorithmic}
\end{algorithm}

Our algorithm may need more than one iteration to calculate the best solution. Following a hill climbing technique, after processing the final index, \algname restarts from the start index and repeats the process to optimize the block edit distance further until either the maximum number of iterations is reached or the program converges, i.e., the result of the iteration is found to be the same as the previous iteration. This recalculation of $N$ further improves accuracy and allows the algorithm to identify a better set of matches. This increases the compute time of the algorithm, but it remains an optional step that substantially enhances the algorithm's accuracy.

The iterations after the first enable the program to improve the results as much as possible with the matches found using the \texttt{Compute-W} function. With each iteration, the accuracy of the result gets closer to the true optimal value; however, it may not necessarily converge to it.

We provide detailed descriptions of the algorithms we use to calculate $N$ in Algorithm~\ref{alg:bes} and Algorithm~\ref{alg:compn}. These algorithms provide step-by-step instructions on making decisions and calculating the block edit distance to find the most optimal operation performed at every index of the source sequence. Using this block edit distance function, the $N$ array can then be constructed using Algorithm~\ref{alg:compn}.

\begin{algorithm}
\caption{Construct N}\label{alg:compn}
\begin{algorithmic}[1]
\Procedure{Construct-N}{$\ell_{min}, \ell_{max}, m, iterations$}
    \State Initialize $N$ with length $m$ and fill with $0$
    \State $BED_{best} \gets \Call{Block-Edit-Distance}{N}$ \Comment{Initial $BED$ = Pairwise Distance} 
    
    \For{$it \gets 1$ to $iterations$}
        \State $BED_{it} \gets BED_{best}$
        \For{$i \gets \ell_{min}$ to $m-1$}  
            \State $k_{best} \gets N[i]$
            \For{\textbf{each} $k$ in $\{0\} \cup [-\ell_{max}$, -$\ell_{min}] \cup [\ell_{min}$, $\ell_{max}]$} \Comment{Skip, Remove and Move Operations}
                \If{$|k| > i$ or $N[i] = k$} 
                    \State \textbf{continue}
                \EndIf
                \State $N[i] \gets k$ 
                \State $BED \gets \Call{Block-Edit-Distance}{N}$
                \If {$BED < BED_{best}$}
                    \State $BED_{best} \gets BED$ 
                    \State $k_{best} \gets k$
                \EndIf
            \EndFor
            \State $N[i] \gets k_{best}$
        \EndFor
        \If{$BED_{it} = BED_{best}$}
            \State \textbf{break} \Comment{Break If Converged}
        \EndIf
    \EndFor
    \State \Return $N$
\EndProcedure
\end{algorithmic}
\end{algorithm}

To maintain the property of disjointness among the blocks in the target sequence, we verify that there is no intersection between blocks before considering any block move operations. To determine whether a block move would violate the disjointness property, the algorithm examines $N$ backward. If any other block move operations intersect with the operation intended to be added, the operation is not taken into consideration.

\subsection{Run time analysis and optimizations}

In this section, we provide some steps to improve the run time of \algname.
    
\subsubsection{Computing W.}

The run time of the na\"ive computation of $W$ (i.e., calculating each individual distance separately), as given above, is $O(m\cdot n \cdot {\ell_{\textit{max}}}^2 \cdot {\ell_{\textit{range}}}^2)$, where $\ell_{\textit{range}} = \ell_{\textit{max}} - \ell_{\textit{min}} + 1$.
However, we can modify the dynamic programming formulation to allow the best match from $T$ for a block in $S$ to be found in a single pass
through using semi-global alignment where the gaps at the start or end of the target sequence are not penalized.

%The modified dynamic programming formulation is that the initial distance along the entirety of $T$ is set to 0, allowing the match to start from anywhere on $T$; and the final row is searched to find its smallest value, allowing the match to end anywhere. 
%Given a block from $S$ of length $\ell$, and the entire sequence $T$ of length $n$, the formulation for constructing the table is as follows: \\
% CEMAL: which table? Is this supposed to be W? If so, why did you write d()?

%\begin{eqnarray*}
%d_{0,0} = 0 &  \\
%d_{0,j} = 0, & ~~ 1 \le j \le n \\
%d_{i, 0}  =  d_{i-1,0} + \sigma_d, & ~~ 1 \le i \le \ell\\
%d_{i, j} =  min\begin{cases} d_{i-1,j} + \sigma_d \\ d_{i,j-1} + \sigma_i \\ d_{i-1, j-1} + %\mu(s_i, t_j) \end{cases} & \begin{array}{c}
%     1 \le i \le \ell  \\
%     1 \le j \le n 
%\end{array}
%\end{eqnarray*}

%$d_{i, j} = min\begin{cases} d_{i-1,j} + C_{Char\_Del} \\ d_{i,j-1} + C_{Char\_Ins} \\ d_{i-1, j-1} + C_{Sub}(a_i, b_j) \end{cases} 1 \le i \le \ell,\ \ 1 \le j \le n \\ \\$

\junk{
\begin{align*}
& d_{0,0} = 0 &  \\
& d_{0,j} = 0, &  1 \le j \le n \\
& d_{i, 0} =  d_{i-1,0} + \sigma_d, &  1 \le i \le \ell\\
& d_{i, j} =  min\begin{cases} d_{i-1,j} + \sigma_d \\ d_{i,j-1} + \sigma_i \\ d_{i-1, j-1} + \mu(s_i, t_j) \end{cases} & \begin{array}{c}
     1 \le i \le \ell  \\
     1 \le j \le n 
\end{array}
\end{align*}

Here $\sigma_d$ is the character deletion penalty, $\sigma_i$ is the character deletion penalty, and $\mu$ refers to the character substitution penalty table.
}

After this modification, the time complexity of computing $W$ becomes $O(m \cdot n \cdot \ell_{\textit{range}} \cdot \ell_{\textit{max}})$, improving the run time by a factor of $O(\ell_{\textit{range}} \cdot \ell_{\textit{max}})$.
Another advantage of this formulation is that the table generated for $S[i..i+\ell_{\textit{max}}-1]$ also contains the necessary information for $S[i..i+\ell-1]$ for every $\ell_{\textit{min}} \le \ell < \ell_{\textit{max}}$. For example, since the last row yields the best match for  $\ell = \ell_{\textit{max}}$, the penultimate row stores the best match for $\ell = \ell_{\textit{max}} - 1$. Thus, only $O(m)$ such tables are needed. This approach reduces the algorithm's run time by another factor of $O(\ell_{\textit{range}})$.
Finally, the time complexity of \texttt{Compute-W} improves to $O(m \cdot n \cdot \ell_{\textit{max}})$.

\subsubsection{Constructing N.}
With the improved time complexity of \texttt{Compute-W}, the greedy construction of $N$ is the bottleneck in the computation of block edit distance. \texttt{Construct-N} uses the \texttt{Block-Edit-Distance} algorithm, which runs in $O(m \cdot n)$ time complexity, and it does this $O(m \cdot \ell_{\text{range}} \cdot \textit{iterations})$ times, where \textit{iterations} is the number of iterations the greedy algorithm runs for. Therefore, the overall time complexity of \texttt{Construct-N} becomes  $O(m^2\cdot n\cdot \ell_{\textit{range}} \cdot \textit{iterations})$.

To optimize the construction of $N$, improvements to calculating the edit distance~\cite{Levenshtein1966} are the most helpful since it dominates the overall run time to construct $N$. Additionally, it is possible to construct $N$ faster by introducing a boundary condition: at index $i$ if the result of block removal of $j_0$ characters is too high $(score > best\_score)$, we do not need to consider block removals of lengths $j_1 \ge j_0 - (score - best\_score - 1)$. 

\subsection{Implementation}
We implemented \algname in C++, using the Edlib~\cite{Sosic2017} library. The source code is available at \url{http://github.com/BilkentCompGen/saber}.

%% file: results.tex
We tested the accuracy of \algname through simulations and portion of the MHC locus in the human genome. \algname does not yet support full genome alignments; therefore we could not use existing genome assemblies such as CHM13 from Telomere-to-Telomere Consortium~\cite{Nurk2022}, or genome variation simulators such as VarSim~\cite{Mu2015}. Thus, we first developed a \textit{rearrangement simulator} that randomly inserts single character mutations and block edits on given sequences.  However, we could not compare accuracy with GR-Aligner, which, to our knowledge, is the only other tool that aims to solve the block edit distance problem with traceback since the code is no longer available. Also, note that genomic rearrangement computation algorithms based on (or extensions of) the HP and DCJ models ~\cite{Bohnenkaemper2020,Shao2015} require sets of genomic markers and do not calculate block edit distance sequence pairs; therefore, they are not directly comparable.

%\clearpage
We used three values when examining the accuracy of \algname:
\begin{itemize}
\item \textbf{Simulated Block Edit Distance:} The block edit distance of the \textit{true} alignment generated by the simulator. We consider it as ground truth, although it may not always be the optimal block edit distance. Denoted as $SBED(S,T)$.
\item \textbf{Calculated Block Edit Distance:} Prediction by \algname. Denoted as $CBED(S,T)$.
\item \textbf{Edit Distance:} Levenshtein Distance between the sequences. Denoted as $ED(S,T)$.
\end{itemize}

Assuming the simulations produce only the optimal block operations, we have  $ED \ge CBED \ge SBED$. Therefore, any \textbf{successfully detected block operation} implies that $CBED$ is ``closer'' to $SBED$. Likewise, an \textbf{undetected block operation} will cause $CBED$ to be higher and closer to $ED$. Thus, in the case of perfect reconstruction of block edit operations $CBED=SBED$, and lack of any block edit operations means $CBED=ED$.

\junk{
We define the \textit{edit error} of each simulation instance as:
\[edit\_error(S,T) = \frac{CBED - SBED}{ED - SBED}\]
and the accuracy as:
\[accuracy = 1 - edit\_error = \frac{ED - CBED}{ED - SBED}\]
}

To formalize empirical correctness estimations of our algorithm, we define the \textit{edit error} and \textit{accuracy} of each simulation instance as follows.
\begin{align*}
edit\_error(S,T) = \frac{CBED - SBED}{ED - SBED}\\
accuracy = 1 - edit\_error = \frac{ED - CBED}{ED - SBED}  
\end{align*}
We then define two more new variables to compute the overall edit error over $k$ simulations: $total\_error$ (sum of the numerators) and $total\_range$ (sum of the denominators). Then, the overall edit error can be calculated as $\frac{total\_error}{total\_range}$. 

We define the \textit{divergence} of a pair of strings $S,T$ by the number of block operations in their alignment. If a sequence $S$ can only allow $b_{max}$ non-overlapping block operations given the minimum and maximum block length, then alignment with $b$ operations has divergence ${b}/{b_{max}}$. Note that $b_{max} = \lfloor m/\ell_{\textit{max}}\rfloor$ ($m=|S|$).

\begin{figure}[!ht]
    \centering
    \includegraphics[width=0.42\textwidth]{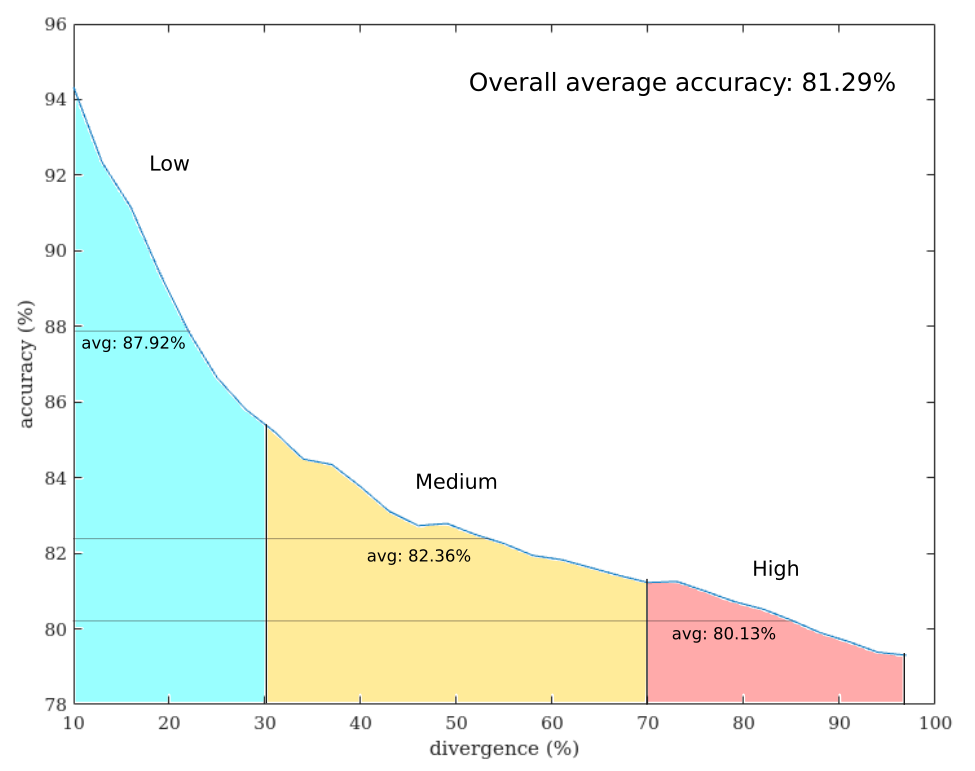}
    \caption{Accuracy of \algname over different testing divergence.}
    \label{fig:results}
\end{figure}

We conducted the simulation experiments for a range of divergence values starting from $10\%$ up to $97\%$ with $3\%$ stepping intervals. We used randomly selected subsequences from the \textit{human reference genome (GRCh38)} as the baseline. For each step, we prepared $15$ simulations using randomly generated sequences of length $800-1200$ and block size range within $20-40$ characters  
and tested our algorithm, allowing at most five iterations during the computation.
In total, we created $450$ simulated $S,T$ pairs. We show the accuracy of the simulations for each divergence interval in Figure~\ref{fig:results}.

We then categorized the divergence levels into three different categories: low $(10\%-30\%)$, medium $(30\%-70\%)$, and high $(\geq 70\%)$. \algname was able to perform well across all divergence level values, although, as expected, its accuracy was inversely proportional with divergence. Briefly,  \algname achieved $87.92$\% average accuracy for the low divergence simulation, where its average accuracy reduced to  $82.36$\% for medium and to $80.13$\% for high divergence sequence pairs. The overall accuracy for the entire simulation was $81.29\%$ (Figure~\ref{fig:results}).

We finally evaluated the time required by \algname. Owing to its time complexity, on average, \algname took 24.8 seconds per sequence pair in the sim. We note that the run time is not dependent on the block edit distance itself; thus, \algname performance is uniform across all divergence levels. Additionally, we note that the current implementation of \algname does not scale well to process strings larger than 5,000 characters.

\paragraph{Aligning the MHC locus.} To provide a test using real sequences, we aligned a portion of the MHC locus from the human reference genome (GRCh38; chr6:28,923,493-28,929,558) to the corresponding sequence from one of the alternate haplotypes (chr6\_GL000251v2\_alt). The reference sequence is 6 Kb, and the alternative sequence is 1.8 Kb. \algname calculated the alignment in 4 minutes. Since this locus only contained large deletions, we compared the block removals predicted by \algname and also by pairwise alignment using Clustal-W~\cite{Thompson1994} with the alignments provided in the UCSC Genome Browser. We observed a better overlap of \algname predictions compared to  Clustal-W when compared against the UCSC-provided alignment  (88.3\% vs 77.2\%; Supplementary Figure~\ref{fig:mhc}).

%% file: supp.tex
\setcounter{table}{0}
\setcounter{figure}{0}
\renewcommand{\thetable}{S\arabic{table}}
\renewcommand\thefigure{S\arabic{figure}}
\setcounter{section}{0}
\renewcommand\thesection{\Alph{section}}

\clearpage
\onecolumn

\section{Supplementary Figures}

\begin{figure}[h]
    \centering
    \includegraphics[width=\textwidth]{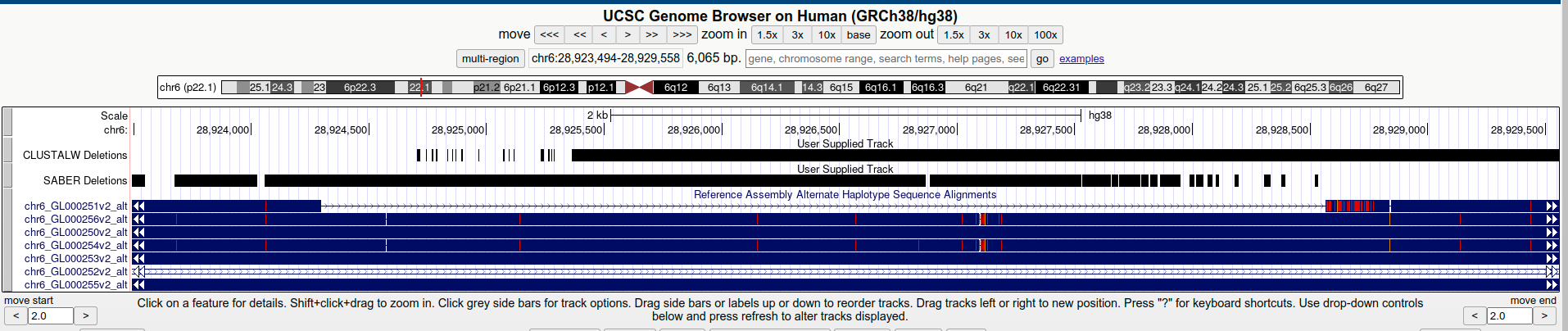}
    \caption{Alignment of a portion of the MHC locus. We selected a 6 Kb region from the human reference genome (GRCh38) that corresponds to the MHC locus, and the corresponding 1.8 Kb portion of an alternative haplotype (chr6\_GL000251v2\_alt). We then used \algname to predict block deletions. We also used Clustal-W~\cite{Thompson1994} to generate optimal pairwise alignment of the two sequences. We show both \algname and Clustal-W predicted deletions as  User Tracks in the UCSC Genome Browser. Deletions predicted by \algname overlap at 88.3\% with the UCSC alignment, where Clustal-W predicted deletions overlap at 77.2\%.}
    \label{fig:mhc}
\end{figure}